\documentclass[submission, Phys]{SciPost}
\begin{document}

\begin{center}{\Large \textbf{
Classical Casimir free energy for two Drude spheres of arbitrary radii: 
A plane-wave approach
}}\end{center}

\begin{center}
Tanja Schoger and Gert-Ludwig Ingold\textsuperscript{*}
\end{center}

\begin{center}
Institut f{\"u}r Physik, Universit{\"a}t Augsburg, 86135 Augsburg, Germany
\\
* gert.ingold@physik.uni-augsburg.de
\end{center}

\begin{center}
\today
\end{center}

\section*{Abstract}
{\bf
We derive an exact analytic expression for the high-temperature limit of the
Casimir interaction between two Drude spheres of arbitrary radii. Specifically,
we determine the Casimir free energy by using the scattering approach in the
plane-wave basis. Within a round-trip expansion, we are led to consider the 
combinatorics of certain partitions of the round trips. The
relation between the Casimir free energy and the capacitance matrix of two
spheres is discussed. Previously known results for the special cases of a
sphere-plane geometry as well as two spheres of equal radii are recovered.
An asymptotic expansion for small distances between the two spheres is determined
and analytical expressions for the coefficients are given. 
}

\vspace{10pt}
\noindent\rule{\textwidth}{1pt}
\tableofcontents\thispagestyle{fancy}
\noindent\rule{\textwidth}{1pt}
\vspace{10pt}

\section{Introduction}

The Casimir effect is often seen as a quantum effect arising from the vacuum
fluctuations of the electromagnetic field between two objects. However, for non-zero 
temperature $T$ also thermal photons with wavelength $\lambda_T =
\hbar c/k_\mathrm{B}T$ contribute to the Casimir force. In fact, for distances
larger than the wavelength $\lambda_T$, the main contribution to the Casimir force
is due to thermal fluctuations. This leads to a finite force even in the 
classical limit of $\hbar\rightarrow 0$ which in view of the definition of the 
thermal wavelength is equivalent to the high-temperature limit $T\rightarrow\infty$. 
The Casimir free energy, which then no longer depends on Planck's constant, is found 
to be linear in temperature. Consequently, the Casimir entropy becomes constant,
thereby revealing the entropic origin of the Casimir effect in the classical
limit \cite{Feinberg2001}. 

Within the scattering approach to the Casimir effect
\cite{LambrechtNetoReynaud2006}, the high-temperature limit amounts to taking
the zero-frequency term of the Matsubara sum. The associated simplification of
the problem has allowed to obtain analytical solutions not only for the
archetypal plane-plane geometry \cite{Sauer1962,Mehra1967} but also for a
scalar field with Dirichlet boundary conditions in the sphere-plane and
sphere-sphere geometry as well as for the electromagnetic field in the
sphere-plane geometry for boundary conditions corresponding to a Drude metal
\cite{BimonteEmig2012}. Even though it was suspected that the extension of the
latter to two spheres of different radii might not be possible
\cite{Bimonte2018} we will see in the following that an analytical expression
for the Casimir free energy in the general setup of two Drude spheres can be
obtained within the scattering approach.

Besides the general theoretical interest in analytical solutions, there is also
practical interest in such an expression. While most Casimir experiments so far
have been carried out using the sphere-plane geometry, the sphere-sphere
geometry has received more attention lately in experiments measuring Casimir
forces \cite{Ether2015,Garrett2018} or addressing colloidal systems
\cite{Elzbieciak-Wodka2014,Ruiz-Cabello2017}. Carrying out the experiment in an
aqueous salt solution offers the opportunity to study the zero-frequency
contribution even outside the high-temperature limit by changing the salt
concentration \cite{Ether2015}.

Furthermore, theoretical results in the high-temperature limit can provide a
crucial ingredient to a semi-analytical approach \cite{Bimonte2018} useful in 
the analysis of experimental data. There, the terms for non-zero Matsubara
frequencies are treated within the derivative expansion. For the zero-frequency
contribution, it is found to be advantageous to employ known exact results
available for the sphere-plane geometry \cite{BimonteEmig2012} and two spheres
with equal radii \cite{Zhao2013}. An exact analytical expression for the setup
of two spheres with arbitrary radii will thus be valuable.

It is common to treat geometries involving one or more spheres within a spherical or 
bispherical multipole expansion. With such approaches the high-temperature limit
of two spheres with different radii has not been explicitly derived so far. However, 
it could have been obtained by combining the results of \cite{BimonteEmig2012} and  
\cite{Fosco2016} together with the capacitance matrix discussed in
Section~\ref{sec:capacitance_matrix}. In \cite{BimonteEmig2012}, bispherical 
coordinates where employed to determine the free energy for two Dirichlet spheres 
while \cite{Fosco2016} applied field theoretical methods to calculate the 
difference in free energy of two spheres for Dirichlet and Drude boundary conditions.

Here, we will take a different approach by working
in the plane-wave basis which has been shown to allow for interesting physical insights
\cite{Spreng2018,Henning2019} as well as an efficient numerical method \cite{Spreng2020}.
Our derivation of the Casimir free energy of two Drude spheres with arbitrary radii will
entirely be based on the plane-wave basis. A round-trip expansion of the scattering of
electromagnetic waves between the two spheres leads to an interesting combinatorial problem
which can be solved. Furthermore, our calculation sheds light on a relation between the
scattering approach to the Casimir effect and a problem of electrostatics.

The paper is organized as follows. Section~\ref{sec:plane-wave-basis} introduces the
scattering approach within the plane-wave basis, where we express the Casimir
free energy as a sum over round trips between the two spheres. In
Section~\ref{sec:scalar_field} we illustrate the basic idea of our approach by
deriving an exact expression for the Casimir free energy of a scalar field.
Our result is found to be dual to the known result \cite{BimonteEmig2012} in the sense
that the free energy is obtained by summing over round trips instead of bispherical multipoles. By
evaluating the spherical monopole contributions in Section~\ref{sec:Drude_spheres} and
subtracting them from the free energy of the scalar field, we obtain as our main result an
exact expression for the Casimir free energy for an electromagnetic field in
the presence of two Drude spheres. It turns out that the monopole contributions can be related
to the capacitance matrix of the sphere-sphere geometry
\cite{Fosco2016,Smythe1950}. Furthermore, we show that our result for the Casimir free energy
agrees with previously obtained expressions for the sphere-plane geometry
\cite{BimonteEmig2012} and two spheres of equal radii \cite{Zhao2013}. Finally,
in Section~\ref{sec:shortdistance}, the short-distance expansion for the general
sphere-sphere geometry is derived with some technical details relegated to the
appendix.

\section{Classical Casimir free energy within the plane-wave basis}\label{sec:plane-wave-basis}

\begin{figure}
\centering
\includegraphics[scale=1]{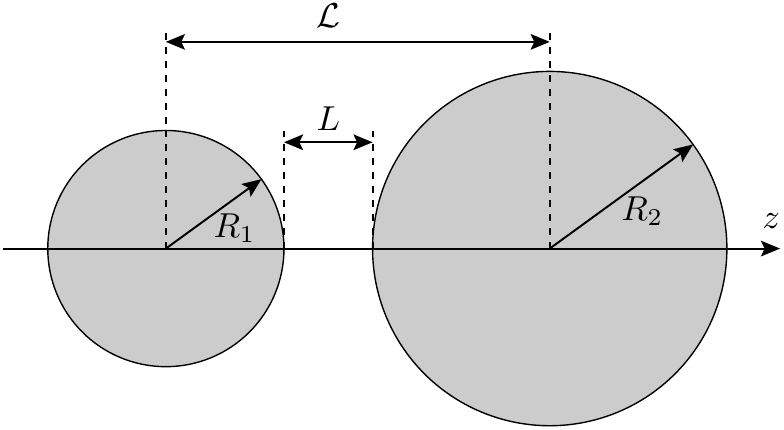}
\caption{Representation of the geometry with two spheres of radii $R_1$ and $R_2$.
	 The distance between the spheres is given by $L$ and $\mathcal{L}$ defines
	 the separation between the sphere centres.}
\label{fig:geometry}
\end{figure}

We start by compiling all ingredients required to evaluate the Casimir free
energy in the high-temperature limit within the scattering approach. The
geometry of our sphere-sphere setup is shown in Fig.~\ref{fig:geometry} where the
two spheres have generally different radii $R_1$ and $R_2$ and are placed at a
centre-to-centre distance $\mathcal{L} = R_1+R_2+L$. $L$ denotes the smallest
distance between the two sphere surfaces. The $z$-axis is chosen to go through
the spheres' centres. Furthermore, the spheres are assumed to be made of a
Drude-type metal with a dielectric function
\begin{equation}
\label{eq:dielectric_function}
\epsilon(i\xi) = 1+ \frac{\omega_\mathrm{P}^2}{\xi(\xi + \gamma)}
\end{equation}
for imaginary frequencies $\omega = i \xi$ with the plasma frequency $\omega_\mathrm{P}$ 
and the relaxation frequency $\gamma$. The dielectric function \eqref{eq:dielectric_function} 
implies a finite dc conductivity $\omega_\mathrm{P}^2/\gamma$. As a consequence, only the 
electric modes contribute to the Casimir energy in the high-temperature limit as we will see below.

Within the scattering approach to the Casimir effect 
\cite{LambrechtNetoReynaud2006}, the free energy is obtained by summation over 
terms containing the round-trip operator $\mathcal{M}$ at the Matsubara frequencies 
$\xi_n = 2\pi n/k_\mathrm{B}T$. Here, $k_\mathrm{B}$ 
and $T$ are the Boltzmann constant and the temperature,
respectively. In the high-temperature limit $L/\lambda_T\gg 1$, only the
zero-frequency term is relevant and the Casimir free energy becomes
\begin{equation}
\label{eq:scattering_approach}
\mathcal{F} = \frac{k_\mathrm{B}T}{2}\mathrm{tr}\log\left[1-\mathcal{M}(\xi = 0)\right]\,. 
\end{equation}
The round-trip operator $\mathcal{M}$ describes one complete round trip of the
electromagnetic waves between the two spheres and is defined as
\begin{equation}
\mathcal{M} = \mathcal{R}_2 \mathcal{T}_{21}\mathcal{R}_1\mathcal{T}_{12}\,. 
\end{equation}
$\mathcal{R}_{1}$ and $\mathcal{R}_{2}$ are the reflection operators for the
two spheres while the operators $\mathcal{T}_{12}$ and $\mathcal{T}_{21}$
describe the translation between the centres of the spheres. In the following,
we omit the argument of the round-trip operator because we will exclusively be
concerned with the zero-frequency case. 

For our purpose, it is convenient to expand the logarithm appearing in
\eqref{eq:scattering_approach} into a Mercator series. The Casimir free energy
then reads
\begin{equation}
\label{eq:mercator}
\mathcal{F} = -\frac{k_\mathrm{B}T}{2} \sum_{r=1}^\infty \frac{\mathrm{tr}\mathcal{M}^r}{r}
\end{equation}
which in physical terms amounts to an expansion in the number $r$ of round trips.

In order to evaluate the trace in \eqref{eq:mercator}, we have to choose a
basis.  While it may appear as natural to use spherical
\cite{CanaguierDurand2010,MaiaNeto2008,Emig2008} or bispherical
\cite{BimonteEmig2012} multipoles, we found it convenient to make use of a
plane-wave basis which has been proven useful lately in the study of the
sphere-sphere geometry \cite{Spreng2020,Spreng2018}.

Specifically, we use the angular spectral representation
\cite{Nieto-Vesperinas2006} consisting of plane waves denoted by
$\vert\mathbf{k}, p, \phi\rangle$. Here, $\mathbf{k}$ refers to the projection
of the wave vector onto the plane perpendicular to the $z$-axis.  The
polarization $p$ can be transverse electric (TE) or transverse magnetic (TM)
with respect to the Fresnel plane spanned by the $z$-axis and the incoming wave vector.
Introducing the Wick rotated $z$-component 
of the wave vector $\kappa$, we obtain from the dispersion relation
\begin{equation}
\kappa = \left(\mathbf{k}^2+\frac{\xi^2}{c^2}\right)^{1/2}\,.
\end{equation}
Since the imaginary frequency $\xi$ is preserved during a round trip, we do not
include it in the parameters characterizing the plane-wave basis. Furthermore,
$\xi=0$ in the high-temperature limit considered here, so that $\kappa=\vert
\mathbf{k}\vert$. Finally, $\phi=\pm$ specifies the direction along the $z$-axis
in which the plane wave decays. $\phi$ changes its sign at each reflection.

In the angular spectral representation, the trace of the $r$-th power of the
round-trip operator in the plane-wave basis can now be expressed as
\begin{equation}
\begin{aligned}
\label{eq:def_trMr}
\mathrm{tr}\mathcal{M}^r &= \sum_{p_1,\ldots, p_{2r}} \int\frac{\mathrm{d}\mathbf{k}_1\ldots
	\mathrm{d}\mathbf{k}_{2r}}{(2\pi)^{4r}} 
\prod_{j=1}^{r} e^{-\kappa_{2j}\mathcal{L}} e^{-\kappa_{2j-1}\mathcal{L}}\\
&\quad
\times \langle \mathbf{k}_{2j+1}, p_{2j+1}, - \vert
\mathcal{R}_2 \vert \mathbf{k}_{2j}, p_{2j}, + \rangle
\langle \mathbf{k}_{2j}, p_{2j}, + \vert
\mathcal{R}_1 \vert \mathbf{k}_{2j-1}, p_{2j-1}, - \rangle\,,
\end{aligned}
\end{equation}
where the indices $2r+1$ and $1$ are identified to account for the trace.
The exponential factors represent the diagonal matrix elements of the two translation
operators covering the distance $\mathcal{L}$ between the centres of the spheres.
This latter choice allows us to make use of the standard reflection operators with
the origin of the reference frame at the spheres' centres.

The expression \eqref{eq:def_trMr} requires the knowledge of the matrix elements
of the reflection operator. We concentrate on the results found in the limit of
vanishing imaginary frequency $\xi$ and refer the reader to \cite{Spreng2018}
for more details. The matrix elements are obtained from the Mie scattering
amplitudes by transforming from the polarization basis referring to the Fresnel
plane to the polarization basis referring to the scattering plane. The Mie
scattering amplitudes can be expressed in terms of a sum over multipoles $\ell$
and consist of the angle functions $\tau_\ell(\cos(\Theta))$ and $\pi_\ell(\cos(\Theta))$
accounting for the scattering geometry and the material-dependent Mie
coefficients $a_\ell$ and $b_\ell$ \cite{BohrenHuffman2004}. For imaginary
frequencies, the scattering angle $\Theta$ is defined through $\cos(\Theta) =
-c^2(\mathbf{k}_j\cdot \mathbf{k}_i + \kappa_j\kappa_i)/\xi^2$. 

The low-frequency behavior of the electric Mie coefficient for spheres made of a
Drude metal is given by $a_\ell\sim \xi^{2\ell+1}$. The magnetic Mie
coefficient $b_\ell$ contains an additional power of $\xi$ and can thus be
neglected with respect to $a_\ell$. The low-frequency behavior of the two angle
functions appearing in the Mie scattering amplitudes is found as
$\tau_\ell(\cos(\Theta))\sim \xi^{-2\ell}$ and $\pi_\ell(\cos(\Theta))\sim\xi^{-2\ell+2}$.
Therefore, in the limit of vanishing $\xi$, only the combination
$a_\ell\tau_\ell$ and thus only the Mie scattering amplitude for waves with
polarization lying in the scattering plane contributes.

In the polarization basis taken with respect to the Fresnel plane, it follows that
in the zero-frequency limit only the matrix element
\begin{equation}
\label{eq:reflectionOperator}
\langle \mathbf{k}_j, \mathrm{TM}, \pm|\mathcal{R}| \mathbf{k}_i, \mathrm{TM}, \mp\rangle 
= \frac{2\pi R}{k_j} \sum_{\ell=1}^\infty
\frac{R^{2\ell}}{(2\ell)!}\left[2k_ik_j \left(1+\cos(\varphi_i-\varphi_j)\right)\right]^\ell
\end{equation}
differs from zero. Here, we have expressed the
transverse wave vector $\mathbf{k}_i$ in polar coordinates through the modulus
$k_i$ and the angle $\varphi_i$. The sum over the multipoles $\ell$ can be carried
out and the non-vanishing reflection matrix becomes
\begin{equation}
\label{eq:reflectionmatrixelement}
\langle \mathbf{k}_j, \mathrm{TM}, \pm|\mathcal{R}| \mathbf{k}_i, \mathrm{TM}, \mp\rangle = \frac{2\pi R}{k_j}
\left\{\cosh\left[2R\sqrt{k_ik_j} \cos\left(\frac{\varphi_i-\varphi_j}{2}\right)\right] -1\right\}\,. 
\end{equation}
Note the subtraction of 1 because of the missing monopole term $\ell=0$ in
(\ref{eq:reflectionOperator}) which distinguishes the electromagnetic from
the scalar case.

After inserting the reflection matrix element
\eqref{eq:reflectionmatrixelement} into the expression \eqref{eq:def_trMr} for
the trace, it is convenient to switch to Cartesian coordinates $x_i =
(k_i\mathcal{L})^{1/2}\cos(\varphi_i/2)$ and $y_i =
(k_i\mathcal{L})^{1/2}\sin(\varphi_i/2)$
\begin{equation}
\begin{aligned}
\label{eq:trMr}
\mathrm{tr}\mathcal{M}^r &= \frac{(\rho_1\rho_2)^r}{\pi^{2r}}
\int \mathrm{d} \mathbf{x} 
\int \mathrm{d} \mathbf{y}  
\prod_{j=1}^r 
e^{-\left(x_{2j}^2 + y_{2j}^2\right)} 
e^{-\left(x_{2j-1}^2 + y_{2j-1}^2\right)}
\\
&\quad\times 
\left[\cosh(\chi_{2j}^{(2)}) - 1\right] 
\left[\cosh(\chi_{2j-1}^{(1)}) - 1\right]\,.
\end{aligned}
\end{equation}
Here, $\rho_n = R_n/\mathcal{L}$ denotes the dimensionless radius of spheres
$n=1,2$ and the argument of the hyperbolic cosines is abbreviated by
$\chi_i^{(n)} = 2\rho_n\left(x_ix_{i+1} + y_i y_{i+1}\right)$. The trace over
the $r$-th power of the round-trip operator is now given by a sum over
$2r$-dimensional Gaussian integrals. After having determined the matrices
associated with the bilinear forms in the exponentials, our main task will be to
evaluate the corresponding determinants.

As already remarked above, the subtraction of 1 in the last two factors in
\eqref{eq:trMr} arises because the monopole term does not contribute in the case of
electromagnetic waves. Including the monopole term amounts to considering the
case of a scalar field with Dirichlet boundary conditions on the spheres.  In
the literature \cite{BimonteEmig2012}, it has been found useful to first
evaluate the scalar case and then to determine the correction corresponding to
the monopole contributions. In the next section, we will thus consider the scalar
case. The plane-wave approach will lead us to an expression for the Casimir
free energy which is equivalent to the known result \cite{BimonteEmig2012}.

\section{Scalar field with Dirichlet boundary conditions}
\label{sec:scalar_field}

According to the discussion in the previous section, the trace over the $r$-th
power of the round-trip operator for a scalar field and two spheres with
Dirichlet (D) boundary conditions can be expressed in the plane-wave basis as
\begin{equation}
\begin{aligned}
\label{eq:trMr_Dirichlet}
\mathrm{tr}\mathcal{M}^r_{(\mathrm{D})} &= \frac{(\rho_1\rho_2)^r}{(2\pi)^{2r}}
\int \mathrm{d} \mathbf{x} 
\int \mathrm{d} \mathbf{y}  
\prod_{j=1}^r 
e^{-\left(x_{2j}^2+ y_{2j}^2\right)} 
e^{-\left(x_{2j-1}^2+ y_{2j-1}^2\right)}
\\
&\quad\times 
\left[e^{\chi_{2j}^{(2)}} + e^{-\chi_{2j}^{(2)}}\right] 
\left[e^{\chi_{2j-1}^{(1)}} + e^{-\chi_{2j-1}^{(1)}}  \right].
\end{aligned}
\end{equation}
Expanding the product, one obtains a sum over $2^{2r}$ Gaussian integrals where
the bilinear form in the exponent can be written with the help of the $2r$-dimensional 
symmetric matrix
\begin{equation}
\label{eq:M_pm}
\mathbf{M}_r^{\pm} = \begin{pmatrix}
1 & \pm\rho_1 & 0 & \ldots  & 0 & \pm\rho_{2} \\ 
\pm\rho_1 & 1 & \pm\rho_2 &   &  &  0 \\ 
0 & \pm\rho_2 & 1 & \ddots &   & \vdots  \\ 
\vdots  &   & \ddots & \ddots &   &  0  \\ 
0 &   &   &   &   & \pm\rho_{1} \\ 
\pm\rho_{2} & 0  &  \ldots &  0 & \pm\rho_{1} & 1
\end{pmatrix}\,,
\end{equation} 
where all combinations of signs appear in the expansion of the product in
\eqref{eq:trMr_Dirichlet}. As far as the determinant is concerned, it only
matters whether the number of minus signs in the upper or lower half of the
matrix is even or odd as indicated by the superscript $+$ or $-$ of
$\mathbf{M}_r^\pm$, respectively.

For a single round trip, $r=1$, the determinant reads
\begin{equation}
\det\mathbf{M}_1^\pm = 2\rho_1\rho_2\left(y\mp1\right)\,,
\end{equation}
where
\begin{equation}
\label{eq:def_y}
y = \frac{1-\rho_1^2-\rho_2^2}{2\rho_1\rho_2} = 1 + \frac{L}{R_\mathrm{eff}}
    +\frac{L^2}{2R_\mathrm{eff}(R_1+R_2)}
\end{equation}
characterizes the geometry of the sphere-sphere arrangement with the effective
radius $R_\mathrm{eff} = R_1R_2/(R_1+R_2)$. For a general number of round trips,
it can be useful to view \eqref{eq:M_pm} as the Hamiltonian matrix of a periodic
tight-binding model and to reexpress the problem in terms of transfer matrices
\cite{Molinari1997,Molinari2008}. One then finds
\begin{equation}
\label{eq:detM0pm}
\det\mathbf{M}_r^{\pm} = 2(\rho_1\rho_2)^r\left[\cosh(r\mu)\mp 1\right]
\end{equation}
with
\begin{equation}
\label{eq:def_mu}
\mu = \mathrm{arcosh}(y).
\end{equation}

We are now in a position to evaluate the Gaussian integrals in \eqref{eq:trMr_Dirichlet}.
Noting that the bilinear form in the exponent is given by $\mathbf{M}_r^+$ and $\mathbf{M}_r^-$
in half of the terms each, we find with \eqref{eq:mercator} the Casimir free energy for a scalar
field and Dirichlet boundary conditions in the high-temperature limit as a sum over
round trips
\begin{equation}
\label{eq:Dirichlet_roundtrip_sum}
\mathcal{F}_{\mathrm{(D})} = -\frac{k_\mathrm{B}T}{2} \sum_{r=1}^\infty
\frac{1}{2r}\frac{\cosh(r\mu)}{\sinh^2(r\mu)}\,.
\end{equation}

Our result \eqref{eq:Dirichlet_roundtrip_sum} can be viewed as a dual representation of 
the earlier result presented in \cite{BimonteEmig2012}. Following the notation introduced
there, we define
\begin{equation}
\label{eq:def_Z}
Z = \exp(-\mu)
\end{equation}
and write the Casimir free energy as
\begin{equation}
\label{eq:Dirichlet_roundtrip_sum_alt}
\mathcal{F}_{\mathrm{(D})} = -\frac{k_\mathrm{B}T}{2}Z\frac{\mathrm{d}}{\mathrm{d}Z} \sum_{r=1}^\infty
\frac{1}{r^2}\frac{Z^r}{1-Z^{2r}}\,.
\end{equation}
Expanding the last factor in a geometric series, the sum over $r$ can be evaluated. With
the help of the Mercator series, one finally obtains
\begin{equation}
\label{eq:Dirichlet_angularmom_sum}
\mathcal{F}_{\mathrm{(D)}} = \frac{k_\mathrm{B}T}{2}\sum_{l=0}^\infty (2l+1) \log(1- Z^{2l+1})\,,
\end{equation}
in agreement with the result derived by means of bispherical coordinates \cite{BimonteEmig2012}. 

Particularly for small distances, the round-trip representation 
\eqref{eq:Dirichlet_roundtrip_sum} may be numerically advantageous as compared to the 
expansion \eqref{eq:Dirichlet_angularmom_sum} because it possesses better convergence 
properties. It is also straightforward to read off the Casimir
free energy within the proximity force approximation by simply retaining the leading order of
the hyperbolic functions. Details of the asymptotic expansion in $\mu$ of the Casimir free energy
will be discussed in section~\ref{sec:shortdistance}.

\section{Electromagnetic case for two general Drude spheres}
\label{sec:Drude_spheres}

\subsection{Monopole contributions in the scalar case}
\label{subsec:roundtrip_partition}

The trace over the $r$-th power of the round-trip operator in the electromagnetic case
differs from the scalar case only by the monopole term $\ell=0$ as one can see by
comparing the corresponding expressions \eqref{eq:trMr} and \eqref{eq:trMr_Dirichlet}.
As already discussed at the end of section 2, does the $-1$ in the brackets of 
\eqref{eq:trMr} account for the subtraction of the monopole term. Hence, when expanding 
the product in \eqref{eq:trMr}, all summands containing at least one of these terms $-1$ 
yield the negative monopole contributions present in the scalar case. 
In the following, we will thus focus on the difference
\begin{equation}
\label{eq:deltar}
\Delta_r = \mathrm{tr}\mathcal{M}^r - \mathrm{tr}\mathcal{M}_{\mathrm{(D)}}^r\,. 
\end{equation}
Already in previous works it was found convenient to study the difference between
the scalar case with Dirichlet boundary conditions and the electromagnetic case for
Drude-type objects \cite{BimonteEmig2012,Fosco2016}.

The difference $\Delta_r$ consists of a sum over Gaussian-type integrals where the
bilinear form in the exponent is represented by a tridiagonal matrix with the
off-diagonal matrix elements arising from the hyperbolic cosines. Whenever in
the expansion of the product in \eqref{eq:trMr} a factor $-1$ appears, the
corresponding pair of off-diagonal matrix elements vanishes.  In contrast to
the matrix \eqref{eq:M_pm} in the scalar case, the matrix representing the
bilinear form in the exponent of the integrand in \eqref{eq:trMr} is now
block-diagonal and can be written as
\begin{equation}
\mathbf{M}_w = \mathrm{diag}\left(
\mathbf{m}^{(t_1)}_{n_1} \mathbf{m}^{(t_2)}_{n_2} 
\mathbf{m}^{(t_3)}_{n_3}\ldots \mathbf{m}^{(t_k)}_{n_k}
\right)\,,
\end{equation}
where $w$ denotes an element of a set $\Pi_{2r,k}$ containing a multiset of tuples
$\{(n_1, t_1), (n_2, t_2),$ $(n_3, t_3),\ldots,  (n_k, t_k)\}$ with $\sum_i n_i = 2r$ for
$r$ round trips. Each block is a symmetric tridiagonal 2-Toeplitz matrix \cite{GoverBarnett1985} 
of the form
\begin{equation}
\label{eq:m_n}
\mathbf{m}_{n}^{(1/2)} = \begin{pmatrix}
1		  & \pm\rho_{1/2} & \cdots   &       & 0 \\
\pm\rho_{1/2} & 1         & \pm\rho_{2/1} &       &   \\
\vdots	  & \pm\rho_{2/1} & 1		  & \ddots    &   \\
          &			  &  \ddots  &\ddots &   \\
0		  &           &          &       & 1
\end{pmatrix}\,,
\end{equation}
where pairs of off-diagonal matrix elements alternate between $\rho_1$ and $\rho_2$ and
each pair can come with an arbitrary sign. Each block is characterized by its size $n$
and the index of the first off-diagonal entry, indicated by the superscript 1 or 2 and 
accounted for by $t_i$ in the multiset $w$. 
As we will see in more detail later, we cannot set $t_i$ to $1$ or $2$ freely.
Rather, its value needs to be compatible with the values of
$n_{i-1}$ and $t_{i-1}$. 

The result of the Gaussian integration will involve the determinant of $\mathbf{M}_w$ which
equals the product of the determinants of the individual blocks. 
For odd dimension, the determinant of the blocks \eqref{eq:m_n} is given by \cite{Gover1994}
\begin{equation}
\label{eq:det_m_odd}
\det \mathbf{m}_{2k+1}^{(1/2)} = (\rho_1\rho_2)^k U_k(y)\,, 
\end{equation}
while for even dimension one has to distinguish between blocks starting with $\rho_1$
or $\rho_2$ on the off-diagonal
\begin{equation}
\label{eq:det_m_even}
\det \mathbf{m}_{2k}^{(1/2)} = (\rho_1\rho_2)^k \left[U_k(y) + \frac{\rho_{2/1}}{\rho_{1/2}} U_{k-1}(y)\right]\,.
\end{equation}
Here, $U_k$ denotes the Chebyshev polynomial of the second kind and order $k$ while $y$ 
has been introduced in \eqref{eq:def_y} and characterizes the geometry of the sphere-sphere
setup. We note that the determinants do not depend on the choice of signs in \eqref{eq:m_n}
so that in view of the Gaussian integration we end up with $2^{n-1}$ equivalent blocks $\mathbf{m}_{n}^{(1/2)}$.

For the monopole contributions \eqref{eq:deltar} we now obtain together with \eqref{eq:trMr}
and \eqref{eq:trMr_Dirichlet}
\begin{equation}
\Delta_r = \frac{(\rho_1\rho_2)^r}{\pi^{2r}} 
\sum_{k=1}^{2r} (-1)^k \sum_{w\in \Pi_{2r, k}}
\int \mathrm{d}\mathbf{x}\, e^{-\mathbf{x}^t \mathbf{M}_{w}\mathbf{x}}
\int \mathrm{d}\mathbf{y}\, e^{-\mathbf{y}^t \mathbf{M}_{w}\mathbf{y}}\,,
\end{equation}
where the number of blocks in $\mathbf{M}_w$ is given by the summation index $k$ and
$\mathbf{x}^t$ denotes the transpose of $\mathbf{x}$. In the derivation, we have taken
into account that a factor $-2$ is associated with each block as one can see by
evaluating the product. Furthermore, we have accounted for the multiplicity
related to the signs in the block matrices mentioned above. Evaluating the
Gaussian integrals, we arrive at
\begin{equation}
\label{eq:sum_blockmatrices}
\Delta_r = (\rho_1\rho_2)^r \sum_{k=1}^{2r} (-1)^k \sum_{w\in \Pi_{2r, k}} \frac{1}{\det \mathbf{M}_w}\,.
\end{equation}

\subsection{Combinatorics of blocks and diagrammatic representation}

The decomposition of the block matrix $\mathbf{M}_w$ into blocks can be
conveniently analyzed in terms of a diagrammatic representation.  For $r$
round trips, we consider a graph consisting of a chain of $2r+1$ nodes
where the last node should be identified with the first one. These nodes
represent the two spheres and are depicted successively in black and white
corresponding to spheres 1 and 2, respectively. Each block
$\mathbf{m}_n^{(1/2)}$ is represented by a black or white line where the colors
refer to the superscripts 1 and 2, respectively. Therefore, the color of a line
equals the color of the node from which the line starts, reading the diagram
from left to right. The length of the line is given by the dimension $n$ of the
corresponding block.

\begin{figure}
\centering
\includegraphics[scale=0.8]{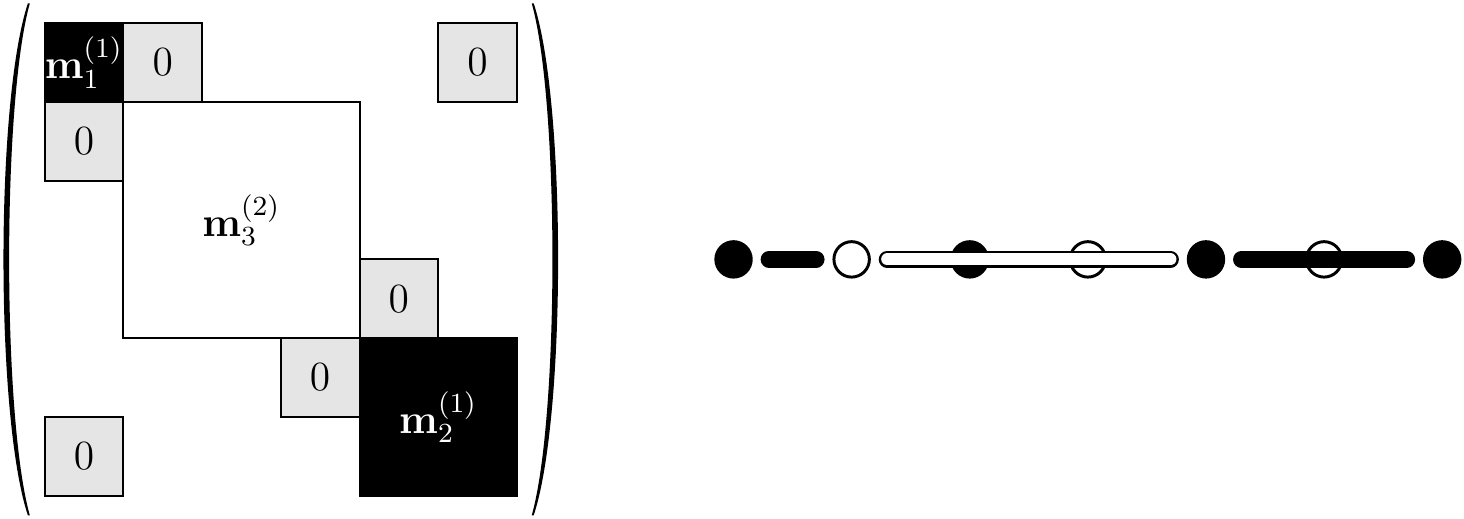}
\caption{An example for a block matrix associated with a term contained in the 
	monopole contributions \eqref{eq:sum_blockmatrices} for $r =2$, $k =3$ is shown on 
	the left together with the corresponding diagrammatic representation depicted on the 
	right-hand side of the figure. In the diagram, the color of a line is determined by the 
	color of the node to its left.}
\label{fig:comparison_with_block_matrix}
\end{figure}

The connection between the block matrix and its diagrammatic representation is
illustrated in Fig.~\ref{fig:comparison_with_block_matrix}. In this example, the
round trip starts on sphere 1 and the block $\mathbf{m}_1^{(1)}$ represents
half a round trip ending on sphere 2. Given the odd dimension of the first
block, the color switches to white. The next block,
$\mathbf{m}_3^{(2)}$, has again an odd dimension and corresponds to one and a
half round trips. In general, it is not required that matrices of odd dimension
follow each other directly. We are now back to sphere 1 and it follows a black
line symbolizing the block $\mathbf{m}_2^{(1)}$, i.e. a single round trip.
This example illustrates why the values of a tuple $(n_i, t_i)$ in the multiset
$w$ depend on the values of the preceding tuple as mentioned earlier.

Disregarding the color for a second, there exists an obvious connection to the
problem of integer partition. It is well known that so-called ordinary Bell polynomials
defined through
\begin{equation}
\label{eq:ordinary_partial_Bell}
\left(\sum_{i=1}^\infty c_i x^{i}\right)^k = 
\sum_{n=k}^\infty \hat B_{n,k}(c_1, c_2, \ldots)x^n
\end{equation}
provide such a partition \cite{Comtet1974}. We will have the opportunity to employ
this relation later when deriving the short-distance behavior of the Casimir free
energy. However, for the following discussion we will need to keep the color because
according to \eqref{eq:det_m_even} the determinant of our blocks for even
dimension depends on the superscript.

To the best of our knowledge, a generalization of the Bell polynomials to our situation
with color does not exist. Fortunately, the dependence of the color of a line on
the length and color of the previous line allows us to express the partitions in
a recursive way instead.

Let us consider $r$ round trips, i.e.\ a chain of length $2r$. We introduce
functions $h_{2r}^{(1)}$ and $h_{2r}^{(2)}$ as sums over the inverse
determinants of all possible block matrices for $r$ round trips starting on
spheres 1 and 2, respectively. These functions will allow us later to express
the monopole contributions $\Delta_r$ as given by \eqref{eq:sum_blockmatrices}.
For convenience, we define abbreviations for the inverse of the determinants of the
blocks with $n\geq1$
\begin{equation}
\label{eq:def_an_bn}
a_n = \frac{1}{\det \mathbf{m}_n^{(1)}}, \quad
b_n = \frac{1}{\det \mathbf{m}_n^{(2)}}
\end{equation}
which will occur in $\Delta_r$. In our diagrams, the coefficients $a_n$ are thus
associated with black lines and the coefficients $b_n$ with white ones.

We can now express the first of the two required functions, which starts on sphere 1, recursively as
\begin{equation}
\label{eq:h1even}
h_{2r}^{(1)}(t) = ta_{2r} - \sum_{n=1}^{r-1} ta_{2n}h_{2(r-n)}^{(1)}(t)
- \sum_{n=1}^{r} ta_{2n-1}h_{2(r-n)+1}^{(2)}(t)\,,
\end{equation}
where the variable $t$ will later serve to determine the number of blocks.
The first term on the right-hand side of \eqref{eq:h1even} accounts for a single
block of maximal size while the other two terms correspond to matrices consisting
of more than one block. The second term arises from a single block of even
dimension followed by a block matrix starting again from sphere 1. The sum runs
over all possible sizes of the first block. The relative minus sign between
the first and the second term is due to the fact that each new block contributes
a minus sign leading to the factor $(-1)^k$ for $k$ blocks in \eqref{eq:sum_blockmatrices}.
The third term differs from the second one in so far as the first block has
an odd dimension so that the remaining part starts on sphere 2.
Instead of $h^{(1)}$ in the second term, we thus have $h^{(2)}$ in the third term.

To close the system of recursive equations, we similarly derive three more equations
\begin{equation}
\label{eq:h1odd}
h_{2r+1}^{(1)}(t) = ta_{2r+1} - \sum_{n=1}^{r} ta_{2n}h_{2(r-n)+1}^{(1)}(t)
- \sum_{n=0}^{r-1} ta_{2n+1}h_{2(r-n)}^{(2)}(t)
\end{equation}
and for graphs starting on sphere 2
\begin{equation}
\label{eq:h2even}
h_{2r}^{(2)}(t) = tb_{2r} - \sum_{n=1}^{r-1} tb_{2n}h_{2(r-n)}^{(2)}(t)
- \sum_{n=1}^{r} tb_{2n-1}h_{2(r-n)+1}^{(1)}(t)
\end{equation}
and
\begin{equation}
\label{eq:h2odd}
h_{2r+1}^{(2)}(t) = tb_{2r+1} - \sum_{n=1}^{r} tb_{2n}h_{2(r-n)+1}^{(2)}(t)
- \sum_{n=0}^{r-1} tb_{2n+1}h_{2(r-n)}^{(1)}(t)\,.
\end{equation}
We note in passing that these recursion relations can be interpreted as the Laplace
expansion of appropriately chosen Hessenberg matrices but we will not make use of
this fact in the following.

\begin{figure}
\centering
\includegraphics[scale=1]{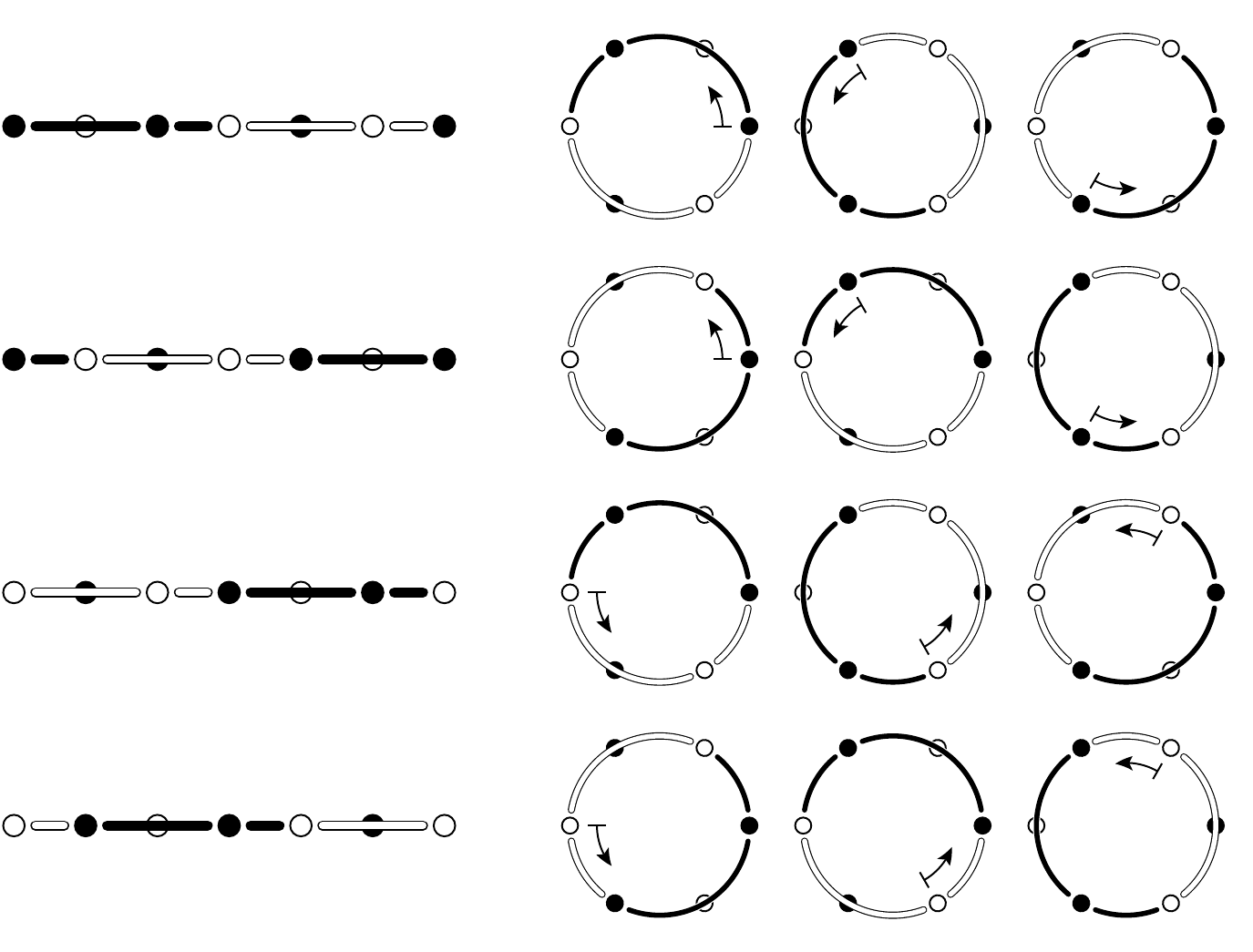}
\caption{Diagrams for the configurations corresponding to
$\mathrm{diag}(\mathbf{m}^{(1)}_2\mathbf{m}^{(1)}_1\mathbf{m}^{(2)}_2\mathbf{m}^{(2)}_1)$,
$\mathrm{diag}(\mathbf{m}^{(1)}_1\mathbf{m}^{(2)}_2\mathbf{m}^{(2)}_1\mathbf{m}^{(1)}_2)$,
$\mathrm{diag}(\mathbf{m}^{(2)}_2\mathbf{m}^{(2)}_1\mathbf{m}^{(1)}_2\mathbf{m}^{(1)}_1)$
and
$\mathrm{diag}(\mathbf{m}^{(2)}_1\mathbf{m}^{(1)}_2\mathbf{m}^{(1)}_1\mathbf{m}^{(2)}_2)$
from top to bottom. The diagrams on the right account for the three possible starting points 
of each partition, when considering a closed chain. The arrows indicate the starting points of the partitions. 
By circularly permuting 
$\mathrm{diag}(\mathbf{m}^{(1)}_2\mathbf{m}^{(1)}_1\mathbf{m}^{(2)}_2\mathbf{m}^{(2)}_1)$,
we reproduce all four configurations. 
On a closed chain the four permutations appear naturally so that each 
diagram occurs four times on the right-hand side.}
\label{fig:multiplicity}
\end{figure}

The sum $h_{2r}^{(1)} + h_{2r}^{(2)}$ accounts for all different kinds of block
matrices occurring for $r$ round trips. However, we have not yet properly
accounted for the multiplicity of the block matrices. So far, we have
considered open-chain diagrams with the starting point chosen at a specific
node for which a pair of off-diagonal elements vanishes. Since the trace at the
origin of \eqref{eq:sum_blockmatrices} implies a closed chain, we can have
several starting points. As Fig.~\ref{fig:multiplicity} demonstrates for $r=3$, a
configuration can start at three different nodes. Specifically, the white line
can start on one of the three white nodes. Since, in general, there are $r$
nodes of one color, we have to multiply the contribution for $r$ round trips by
a factor of $r$.

By proceeding as just described, we include all circular permutations of the
blocks, but the functions $h_{2r}^{(1,2)}$ already include all non-equal
circular permutations. Hence, to avoid double counting, we have to remove the
$k$ cyclic permutations of a partition in $k$ blocks by dividing the
contribution arising from $k$ blocks by $k$. Fig.~\ref{fig:multiplicity}
illustrates a non-trivial example. On the left, we present the diagrams of the partition
$\mathrm{diag}(\mathbf{m}^{(1)}_2\mathbf{m}^{(1)}_1\mathbf{m}^{(2)}_2\mathbf{m}^{(2)}_1)$
and all its circular permutations. As demonstrated by the graphs on the right, 
there are $r=3$ possible ways of choosing a starting point for each partition. However, 
as one can see, each diagram appears four times. Hence, by dividing 
by four, the correct number of block matrices is obtained. 

We now make use of the parameter $t$ introduced
in equations \eqref{eq:h1even}--\eqref{eq:h2odd} which ensures that contributions
arising from $k$ blocks come with a factor $t^k$. The monopole contributions for 
a given number of round trips \eqref{eq:sum_blockmatrices} can thus be written as
\begin{equation}
\label{eq:combinatorics}
\Delta_r = - (\rho_1\rho_2)^r r \int_0^1 \mathrm{d}t \,
\frac{h_{2r}^{(1)}(t) + h_{2r}^{(2)}(t)}{t}\,. 
\end{equation}
The negative sign arises due to our definition of $h_{2r}^{(1,2)}$, where all partitions
in odd numbers of blocks occur with a positive sign compared to those with even numbers. 

So far, we have considered the monopole contributions for a given number of
round trips. The full correction of the Casimir free energy with respect to
the result \eqref{eq:Dirichlet_roundtrip_sum} or
\eqref{eq:Dirichlet_angularmom_sum} for the scalar case is obtained by summation
over all numbers $r$ of round trips, which we will carry out in the following section. 

\subsection{Monopole correction to the classical Casimir free energy}

According to the round-trip expansion \eqref{eq:mercator} of the Casimir free energy,
the monopole contributions are determined by
\begin{equation}
\label{eq:def_Delta}
\Delta = \mathcal{F}-\mathcal{F}_{(\mathrm{D})}
       = -\frac{k_\mathrm{B}T}{2}\sum_{r=1}^\infty \frac{\Delta_r}{r}\,. 
\end{equation}
Strictly speaking, $\Delta$ equals the negative monopole contributions, but for simplicity
we will continue to refer to this quantity as monopole contributions.

After inserting \eqref{eq:combinatorics} and interchanging summation and integration, it
is convenient to introduce the generating functions for the inverse block-matrix determinants
\begin{equation}
\label{eq:def_generating_func}
H^{(1,2)}(x;t) = \sum_{n=1}^\infty h_n^{(1,2)}(t) x^n
= H^{(1,2)}_\mathrm{e}(x;t)+ H^{(1,2)}_\mathrm{o}(x;t)\,.
\end{equation}
In the second equality, we decompose the sum into contributions from even (e) and
odd (o) powers of $x$. Hence, the monopole term yields
\begin{equation}
\label{eq:Delta_integralrepr}
\Delta = \frac{k_\mathrm{B}T}{2}\int_0^1\mathrm{d}t\,
\frac{H_\mathrm{e}^{(1)}(\sqrt{\rho_1\rho_2};t)
	+H_\mathrm{e}^{(2)}(\sqrt{\rho_1\rho_2};t)}{t}\,.
\end{equation}
Correspondingly, we introduce the generating functions for the inverse determinants of
individual blocks \eqref{eq:def_an_bn}
\begin{equation}
\label{eq:ax}
A(x) = \sum_{n=1}^\infty a_n x^n = A_\mathrm{e}(x) + A_\mathrm{o}(x)
\end{equation}
and
\begin{equation}
\label{eq:bx}
B(x) = \sum_{n=1}^\infty b_n x^n = B_\mathrm{e}(x) + B_\mathrm{o}(x)\,,
\end{equation}
where we apply the same decomposition of the functions as in \eqref{eq:def_generating_func}.

The generating functions $H_\mathrm{e}^{(1,2)}$ can be determined by summing over the recurrence 
relations \eqref{eq:h1even}--\eqref{eq:h2odd} and we find 
\begin{equation}
H^{(1)}_\mathrm{e}(x;t) = t\frac{A_\mathrm{e}(x) + tA_\mathrm{e}(x)B_\mathrm{e}(x)
-t  A_\mathrm{o}(x)B_\mathrm{o}(x)}{(1 + tA_\mathrm{e}(x))(1+ tB_\mathrm{e}(x)) 
 -t^2  A_\mathrm{o}(x)B_\mathrm{o}(x)}
\end{equation}
with an analogous expression for $H^{(2)}_\mathrm{e}$ where the functions
$A$ and $B$ are interchanged.
The sum of both even functions yields
\begin{equation}
H^{(1)}_\mathrm{e}(x;t) + H^{(2)}_\mathrm{e}(x;t) =
t\frac{A_\mathrm{e}(x) + B_\mathrm{e}(x)+ 2tA_\mathrm{e}(x)B_\mathrm{e}(x)
- 2tA_\mathrm{o}(x)B_\mathrm{o}(x)}{(1 + tA_\mathrm{e}(x))(1+ tB_\mathrm{e}(x)) 
 -t^2  A_\mathrm{o}(x)B_\mathrm{o}(x)}\,. 
\end{equation}
Noting that the numerator equals the derivative with respect to $t$ of the
denominator, it is straightforward to evaluate the integral in
\eqref{eq:Delta_integralrepr} and we find
\begin{equation}
\label{eq:monopole_contributions}
\Delta = \frac{k_\mathrm{B}T}{2}\log\big[\left(1+ A_\mathrm{e}(\sqrt{\rho_1\rho_2})\right)\left(1+ B_\mathrm{e}(\sqrt{\rho_1\rho_2})\right)
- A_\mathrm{o}(\sqrt{\rho_1\rho_2}) B_\mathrm{o}(\sqrt{\rho_1\rho_2})\big]
\end{equation}
with
\begin{align}
A_\mathrm{e}(\sqrt{\rho_1\rho_2}) &= \sum_{n=1}^\infty \frac{1}{U_n(y) + \alpha U_{n-1}(y)}
\label{eq:a_e}\\
B_\mathrm{e}(\sqrt{\rho_1\rho_2}) &= \sum_{n=1}^\infty \frac{1}{U_n(y) + \beta U_{n-1}(y)}
\label{eq:b_e}\\
A_\mathrm{o}(\sqrt{\rho_1\rho_2}) &=
\sqrt{\rho_1\rho_2}\sum_{n=0}^\infty \frac{1}{U_{n}(y)} 
= B_\mathrm{o}(\sqrt{\rho_1\rho_2})\,,
\label{eq:a_ob_o}
\end{align}
where $\alpha=R_2/R_1$ and $\beta = R_1/R_2$ take the ratios of the sphere radii 
into account and $y$ is defined in \eqref{eq:def_y}. The expressions in \eqref{eq:a_e}--\eqref{eq:a_ob_o}
are obtained by inserting \eqref{eq:def_an_bn} together with \eqref{eq:det_m_odd} and \eqref{eq:det_m_even}
into \eqref{eq:ax} and \eqref{eq:bx}.

It is instructive to convince oneself that indeed all partitions of round trips
are contained in \eqref{eq:monopole_contributions} by expanding the logarithm as
\begin{equation}
\begin{aligned}
\Delta = -\frac{k_\mathrm{B}T}{2}& \left\{ 
\sum_{k=1}^\infty \frac{(-1)^k}{k} 
\left(A^k_\mathrm{e} + B^k_\mathrm{e} \right) \right. \\
&\quad \left. + \sum_{n=1}^\infty \frac{1}{n}
\left[\sum_{l=0}^\infty (-A_\mathrm{e})^l A_\mathrm{o} 
\sum_{m=0}^\infty (-B_\mathrm{e})^m B_\mathrm{o} \right]^n 
\right\}\,.
\end{aligned}
\end{equation}
The first sum accounts for arbitrary repetitions of full round trips starting
either on sphere~1 or on sphere~2 as represented by $A_\mathrm{e}$ or
$B_\mathrm{e}$, respectively. Expressions containing both $A_\mathrm{e}$ and
$B_\mathrm{e}$ can only arise if half round trips represented by $A_\mathrm{o}$
and $B_\mathrm{o}$ occur as is the case in the second term. Reading this term
from left to right, it can clearly be seen that half a round trip induces a
change between full round trips starting on sphere~1 and on sphere~2.  The
number of factors $(-1)$ correctly reflects the number of blocks in the
matrices $\mathbf{M}_w$.

\subsection{Relation to the capacitance matrix}
\label{sec:capacitance_matrix}

It appears that the result \eqref{eq:monopole_contributions} was so far not
known in the Casimir community. Nevertheless, it can be obtained by combining
results from the literature, a fact which we only became aware of after the
work presented here had been carried out. As was shown by Fosco
\textit{et al.} \cite{Fosco2016}, the difference between the Casimir free
energy of objects made of Drude metals and the Casimir free energy for
a scalar field with Dirichlet boundary conditions is related to the capacitance
matrix $\mathbf{C}$ of the arrangement of conductors. For the special case of two conductors,
\cite{Fosco2016} found\footnote{Note that here we adopt the choice of units of \cite{Fosco2016}. 
Furthermore, their quantity $\Delta F$ equals $-\Delta$.}
\begin{equation}
\label{eq:Delta_Fosco}
\Delta = \frac{T}{2}\log\left[\mathrm{det}(\mathbf{C})T^2\right]\,.
\end{equation}

Even though this was not mentioned in \cite{Fosco2016}, the
capacitance matrix elements of two conducting spheres of arbitrary radii
were already known to Maxwell \cite{Maxwell1873}. Following the more modern notation in
\cite{Smythe1950}, the capacitance coefficients can be expressed as
\begin{align}
c_{11} &= R_1(1+B_\mathrm{e}(\sqrt{\rho_1\rho_2}))\\
c_{22} &= R_2(1+A_\mathrm{e}(\sqrt{\rho_1\rho_2}))\\
c_{12} = c_{21} &=  -\sqrt{R_1R_2}A_\mathrm{o}(\sqrt{\rho_1\rho_2})\,.
\end{align}
Comparing these coefficients and \eqref{eq:Delta_Fosco} with our result 
\eqref{eq:monopole_contributions} connects the capacitance coefficients 
to the scattering of electromagnetic waves in the static limit. It thus 
highlights the relation between our round-trip description and the 
method of image charges used by \cite{Maxwell1873} to obtain the 
capacitance coefficients.

We remark that the general result \eqref{eq:Delta_Fosco} and our result
\eqref{eq:monopole_contributions} differ by a factor $R_1R_2T^2$ in the logarithm.
While this factor would be irrelevant for the Casimir force, it makes a
difference for the Casimir entropy. Its origin can be traced back to the
different handling of the Casimir free energy of the individual objects
\cite{BalianDuplantier1978,Feinberg2001}. While the scattering approach
does not contain the free energy of the spheres at an infinite distance,
this contribution is present in \cite{Fosco2016}. For \eqref{eq:monopole_contributions},
the entropy in the high-temperature limit becomes a constant as expected
\cite{Feinberg2001}. 

\subsection{Casimir free energy for two Drude spheres of general radii and limiting cases}

According to \eqref{eq:def_Delta}, the sum of the expressions
\eqref{eq:Dirichlet_roundtrip_sum} and \eqref{eq:monopole_contributions} gives the Casimir
free energy for two Drude spheres of arbitrary radii and thus constitutes the main result
of this paper. Instead of reproducing the two expressions here, it is useful to resum
the result as we did in Section~\ref{sec:scalar_field} for the scalar case and to express it in
terms of the variable $Z$ introduced in \eqref{eq:def_Z}. Noting that the Chebyshev
polynomials of the second kind appearing in equations \eqref{eq:a_e}--\eqref{eq:a_ob_o}
can be written as
\begin{equation}
\label{eq:chebyshev_Z}
U_n(y) = \frac{Z^{-(n+1)} - Z^{(n+1)}}{Z^{-1} - Z}\,,
\end{equation}
we obtain the classical Casimir free energy for two Drude spheres as
\begin{equation}
\begin{aligned}
\label{eq:freeenergy_Z}
\mathcal{F} = \frac{k_\mathrm{B}T}{2}&\left\{\sum_{l=0}^\infty (2l+1) \log(1- Z^{2l+1})\right. \\
&\quad +\log\left[ \left(1+ \frac{1-g_\alpha(Z)^2}{g_\alpha(Z)}
\sum_{l=0}^\infty \frac{(Zg_\alpha(Z))^{2l+1}}{1-Z^{2l+1}}
\right)\right.  \\
& \hspace{6em} \times \left(1+ \frac{1-g_\beta(Z)^2}{g_\beta(Z)}
\sum_{l=0}^\infty \frac{(Zg_\beta(Z))^{2l+1}}{1-Z^{2l+1}}
\right) \\
& \left.\left.
	\hspace{4em}-\frac{(1-g_\alpha(Z)^2)(1-g_\beta(Z)^2)}{Z}
\left(\sum_{l=0}^\infty \frac{Z^{2l+1}}{1-Z^{2l+1}}
\right)^2\right]\right\}\,,
\end{aligned}
\end{equation}
where we have introduced the function
\begin{equation}
\label{eq:def_g}
g_\alpha(Z) = \left(\frac{Z^2+\alpha Z}{1+\alpha Z}\right)^{1/2}
\end{equation}
and correspondingly for $g_\beta(Z)$, where $\alpha$ and $\beta=1/\alpha$ are the ratios
of sphere radii as defined below \eqref{eq:a_ob_o}.

We obtain the limit of a sphere of radius $R$ in front of a plane by setting $R_2=R$ and
letting $R_1$ go to infinity. Then, $g_\beta = 1$ and $B_\mathrm{e}$ and $B_\mathrm{o}$ vanish
because there is no second sphere were the electromagnetic waves could be scattered.
Since the functional dependence of the scalar part of \eqref{eq:freeenergy_Z}, i.e. the first
sum, is not affected, we focus on the monopole contributions for which we obtain
\begin{equation}
\begin{aligned}
\label{eq:Delta_plane_sphere}
\Delta^{(R_1\to\infty)} &= \frac{k_\mathrm{B}T}{2}\log (1+ A_\mathrm{e})\\
&= \frac{k_\mathrm{B}T}{2} 
\log\left[1+ (1-Z^2) \sum_{l=0}^\infty \frac{Z^{4l+1}}{1-Z^{2l+1}}\right]\,,
\end{aligned}
\end{equation}
where $Z$ depends only on the aspect ratio $\epsilon = L/R$ through
\begin{equation}
Z = 1+ \epsilon - \sqrt{\epsilon(2+\epsilon)}\,.
\end{equation}
By some minor transformations, one can convince oneself, that \eqref{eq:Delta_plane_sphere}
agrees with the result found earlier by Bimonte and Emig \cite{BimonteEmig2012}. 

Similarly, we obtain the Casimir free energy for two Drude spheres of equal radii by
setting $R_1=R_2=R$ so that $g_\alpha=g_\beta=Z^{1/2}=Y$. In this case, the scattering
at the two spheres cannot be distinguished and we have $A_\mathrm{e}=B_\mathrm{e}$ and
$A_\mathrm{o}=B_\mathrm{o}$. The monopole contributions then read
\begin{equation}
\begin{aligned}
\Delta^{(R_1 = R_2)} &= \frac{k_\mathrm{B}T}{2}
\left[\log(1+A_\mathrm{e}+A_\mathrm{o}) + \log(1+A_\mathrm{e}-A_\mathrm{o})\right]\\
\label{eq:Delta_equal_spheres}
&= \frac{k_\mathrm{B}T}{2}\left[
\log\left( 1- \sum_{l=1}^\infty \frac{(1-Y^2)(1-Y^{2l})Y^{2l+1}}{1-Y^{2l+1}}
\right)\right.\\
&\qquad\qquad+ \log\left(1+ \sum_{l=1}^\infty \frac{(1-Y^2)(1-Y^{2l})Y^{2l+1}}{1+Y^{2l+1}}\right)
\\
&\qquad\qquad \left.-\log(1-Y^2)\vphantom{\frac{Y^{2l+1}}{Y^{2l+1}}}\right],
\end{aligned}
\end{equation}
where the parameter $Y$ is a function of the aspect ratio $\delta = L/2R$
\begin{equation}
Y = 1+ \delta - \sqrt{\delta(2+\delta)}.
\end{equation}
The result \eqref{eq:Delta_equal_spheres} leads to the same Casimir free energy 
as obtained earlier by using the transformation optics approach \cite{Zhao2013}.

\section{Short-distance expansion}\label{sec:shortdistance}

In experiments, the closest distance $L$ between the two spheres is typically
small compared to the radii $R_1$ and $R_2$. Therefore, we will now determine a
short-distance expansion of the Casimir free energy \eqref{eq:freeenergy_Z} by
separately considering the scalar contribution $\mathcal{F}_\mathrm{(D)}$ and
the monopole contributions $\Delta$. The leading-order term will
correspond to the proximity-force approximation whose validity can be assessed
by the higher-order terms.

In the following, we make use of the fact that $\mathcal{F}_\mathrm{(D)}$ and
$\Delta$ can be expressed in terms of the F-series introduced by Garvin
\cite{Garvin1936} as a generalization of the Lambert series. With a choice of
coefficients appropriate for our situation, we introduce
\begin{equation}
\label{eq:def_Lambert}
\mathcal{L}_q(s, x) = \sum_{k=1}^\infty \frac{k^s q^{kx}}{1- q^k}\,.
\end{equation} 
Here, we follow the notation used by Banerjee and Wilkerson who provide
an asymptotic expansion of this series around $q=1$ \cite{BanerjeeWilkerson2017}.

We start by expanding the Casimir free energy for two Dirichlet spheres. Its
representation \eqref{eq:Dirichlet_roundtrip_sum_alt} can be expressed in terms
of the series \eqref{eq:def_Lambert} as
\begin{equation}
\label{eq:freeenergy_lambert}
\mathcal{F}_{(\mathrm{D})} =  -\frac{k_\mathrm{B}T}{2}Z
\frac{\mathrm{d}}{\mathrm{d}Z}\left[
\mathcal{L}_Z(-2, 1) - \mathcal{L}_{Z^2}(-2, 1)\right]\,.
\end{equation}
For small distances $L\ll R_1, R_2$, the variable $Z= \exp(-\mu)$ is close to unity and $\mu$ 
defined in \eqref{eq:def_mu} is small. In the following, we will use $\mu$ as our expansion
variable.

Making use of the asymptotic expansion of the generalized Lambert series around $q=1$
stated in theorem 2.2 of \cite{BanerjeeWilkerson2017}, we obtain from
\eqref{eq:freeenergy_lambert} for the Casimir free energy in the scalar case
with Dirichlet boundary conditions
\begin{equation}
\begin{aligned}
\mathcal{F}_{(\mathrm{D})} &= \frac{k_\mathrm{B}T}{2}\left[-\frac{\zeta(3)}{2\mu^2}
	+\frac{1}{12}\log(\mu)+\frac{1}{12}-\log(A)+\frac{1}{6}\log(2)\right.\\
	&\qquad\qquad \left.+\sum_{n=1}^\infty\frac{2n+1}{2n}\frac{B_{2n}B_{2n+2}}{(2n+2)!}
	       \left(2^{2n+1}-1\right)\mu^{2n}\right]
\end{aligned}
\end{equation}
with Glaisher's constant $A=1.28242\ldots$ and the Bernoulli numbers $B_k$
\cite{dlmf}. The first term corresponds to the high-temperature result of the
proximity-force approximation with the specific value of the Riemann zeta
function $\zeta(3)=1.20205\ldots$ The terms up to order $\mu^4$ were already
given in \cite{BimonteEmig2012} and are consistent with our result.

Now we turn to the short-distance expansion of the monopole term
as given by the second term of \eqref{eq:freeenergy_Z} where the argument
of the logarithm can again be expressed in terms of a generalized Lambert
series \eqref{eq:def_Lambert}. Readers not interested in the technical details
of the derivation will find the final result in \eqref{eq:short_distance}.

We bring the sums in the monopole contributions $\Delta$ into the form of a generalized
Lambert series by writing the function \eqref{eq:def_g} as
\begin{equation}
g_\alpha = Z^{1/2 + v(\mu)}\,,
\end{equation}
where
\begin{equation}
\label{eq:def_v}
v(\mu) = \frac{1}{2}-\frac{1}{2\mu}
\left[\log(1+\alpha e^{\mu})-\log(1+\alpha e^{-\mu})\right]\,. 
\end{equation}
Replacing $\alpha$ by $\beta$ simply changes the sign of this function, so that
$g_\beta=Z^{1/2-v}$. For our purpose, we need to expand $v(\mu)$ into a Taylor
series
\begin{equation}
\label{eq:taylor_v}
v(\mu) = \sum_{n=0}^\infty v_{n} \mu^{2n}\,.
\end{equation}
The coefficients for $n>0$ can be expressed in terms of
\begin{equation}
\label{eq:def_v0}
v_{0} = \frac{1}{2} \frac{R_1-R_2}{R_1+R_2}
\end{equation}
as
\begin{equation}
\label{eq:def_vn}
v_n = \frac{1}{(2n+1)!}\sum\limits_{k=0}^{2n} k! 
S(2n+1, k+1)\left(v_{0} -1/2\right)^{k+1}\,,
\end{equation}
where $S(n,k)$ are the Stirling numbers of the second kind. 

Before making use of the generalized Lambert series, it is convenient
to introduce a notation for the prefactors and sums appearing in the
argument of the second logarithmic term in \eqref{eq:freeenergy_Z}.
By defining
\begin{equation}
\label{eq:def_Jc}
J(c) = \frac{1-Z^{2(c-1)}}{Z^{c-1}} \quad
\end{equation}
and
\begin{equation}
\label{eq:def_Ic}
I(c) = \sum_{l=0}^\infty \frac{Z^{c(2l+1)}}{1-Z^{2l+1}}\,, 
\end{equation}
the monopole contributions \eqref{eq:monopole_contributions} can be
brought into the form
\begin{equation}
\label{eq:Delta_IJ}
\Delta = \frac{k_\mathrm{B}T}{2}\log \left[
(1+ J^{(+)}I^{(+)})(1 + J^{(-)}I^{(-)})
- J^{(+)}J^{(-)}I^2(1)\right]\,,
\end{equation}
where we introduced the abbreviations $J^{(\pm)} = J(3/2 \pm v)$ and
$I^{(\pm)} = I(3/2 \pm v)$. For the further analysis, we now express the series
\eqref{eq:def_Ic} in terms of the generalized Lambert series \eqref{eq:def_Lambert} as
\begin{equation}
I(c) = \mathcal{L}_Z(0, c) - \mathcal{L}_{Z^2}(0, c)\,.  
\end{equation}
Applying the asymptotic expansion of the generalized Lambert series
\cite{BanerjeeWilkerson2017}, we obtain 
\begin{equation}
\label{eq:expansion_Ic}
I(c) = \frac{1}{2\mu}\left[-\log\left(\frac{\mu}{2}\right) - \psi(c) + 
\sum_{n=1}^\infty \frac{B_{2n}(2^{2n}-2)}{2n(2n)!}B_{2n}(c)\mu^{2n}\right]
\end{equation}
with the digamma function $\psi(c)$ and the Bernoulli polynomial $B_{2n}(c)$
\cite{dlmf}.  For $c=1$, the functions $\psi(c)$ and $B_{2n}(c)$ are given by
the negative Euler-Mascheroni constant, $-\gamma = -0.57721\ldots$ and the
Bernoulli numbers $B_{2n}$, respectively.

Before proceeding with our calculation, we note that a short-distance expansion
of the capacitance coefficients for two general spheres has already been
carried out in \cite{Lekner2011} and \cite{Banerjee2019}, where the latter one
also applied the asymptotic expansion of the generalized Lambert series.
Besides using a different definition of the dimensionless capacitance
coefficients and geometric parameters, we determine, in contrast to previous
work, a complete expansion of the functions $I(c)$ as well as $J(c)$ in powers
of $\mu$. In the definition of our geometric parameters, we follow the notation
common in the Casimir community which also simplifies to obtain the limits
of equal spheres and of the sphere-plane geometry.

In order to obtain a complete expansion of the argument of the logarithm in
\eqref{eq:Delta_IJ} in powers of $\mu$, we need to account for the fact that
$I^{(\pm)}$ and $J^{(\pm)}$ depend on $\mu$ through $c = 3/2\pm v(\mu)$. By
making use of the Taylor series for $v$ given in \eqref{eq:taylor_v} with the
coefficients \eqref{eq:def_v0} and \eqref{eq:def_vn}, one immediately obtains a
corresponding Taylor series for $c(\mu)$ which is required to determine the
Taylor series in powers of $\mu$ for the digamma function $\psi(c(\mu))$ and
the Bernoulli polynomial $B_{2n}(c(\mu))$ appearing in \eqref{eq:expansion_Ic}. 

For $\mu \leq 1$, a condition fulfilled in the small-distance limit, and
with the help of \eqref{eq:ordinary_partial_Bell}, the digamma function can be
expanded as
\begin{equation}
\label{eq:expansion_digamma}
\psi(c(\mu)) = \psi(c_0) + \sum_{m=1}^\infty 
\sum_{n=1}^m \frac{\psi^{(n)}(c_0)}{n!} 
\hat B_{m,n}(c_1, c_2,\ldots) \mu^{2m}\,, 
\end{equation}
where $\psi^{(n)}(c_0)$ denotes the polygamma function \cite{dlmf} and
$\hat B_{m,n}(c_1, c_2,\ldots)$ are partial ordinary Bell polynomials.
Correspondingly, the expansion of the Bernoulli polynomials yields
\begin{equation}
\label{eq:expansion_Bernoulli}
B_{2n}(c(\mu)) = 
\frac{1}{\mu^{2n}}\sum_{k=0}^{2n} \binom{2n}{k}B_k 
\sum_{m= n}^\infty \hat B_{m+n-k, 2n-k}(c_0, c_1, \ldots) \mu^{2m}\,.
\end{equation}
Note that the arguments of the Bell polynomials in \eqref{eq:expansion_digamma} and
\eqref{eq:expansion_Bernoulli} differ.

Inserting the expansions from above into \eqref{eq:expansion_Ic}, we find the
series expansion
\begin{equation}
\label{eq:taylor_Ic}
I(c) = \frac{1}{2\mu}\sum_{m= 0}^\infty I_m(c)\mu^{2m}
\end{equation}
with the coefficients
\begin{equation}
\label{eq:def_I0c}
I_0(c) = -\log(\mu/2) - \psi(c_0)
\end{equation}
and for $m>0$
\begin{equation}
\begin{aligned}
\label{eq:def_Imc}
I_m(c) = 
\sum_{n=1}^m &\left[\frac{B_{2n}(2^{2n} -2)}{2n}
	\sum_{k =-n}^n \frac{B_{n+k}\hat B_{m-k,n-k}(c_0, c_1, \ldots)}{(n-k)!(n+k)!}\right.\\
	& \quad \left. 
	- \frac{\psi^{(n)}(c_0)}{n!} 
\hat B_{m,n}(c_1, c_2,\ldots)\right]\,.
\end{aligned}
\end{equation}
The prefactor \eqref{eq:def_Jc} can be expanded correspondingly and we obtain
\begin{equation}
\label{eq:expansion_Jc}
J(c) = 2\mu \sum_{m=0}^\infty J_m(c) \mu^{2m}
\end{equation}
with the coefficients
\begin{equation}
\label{eq:def_Jmc}
J_m(c)=
\sum_{n=0}^m
\frac{\hat B_{m+n+1, 2n+1}(c_0-1, c_1, c_2,\ldots)}{(2n+1)!}\,. 
\end{equation}

By means of \eqref{eq:taylor_Ic} and \eqref{eq:def_Jmc} one can derive a systematic
expansion of the monopole contributions \eqref{eq:Delta_IJ} for small distances. The optimal
cut-off for this asymptotic series is discussed in Ref.~\cite{Banerjee2019}.
However, even the calculation of the terms up to order $\mu^4$ involves a 
decent amount of algebra which we relegate to Appendix~\ref{sec:appendix_short_distance}.
Expanding the result \eqref{eq:short_distance_appendix} in a Mercator series finally yields 
\begin{equation}
\begin{aligned}
\label{eq:short_distance}
\Delta \approx \frac{k_\mathrm{B}T}{2} &\left\{ 
\log\left[\epsilon_0(\gamma -\log(\mu/2)) + \delta_0\right]
+\frac{1}{6}\frac{\epsilon_1(\gamma -\log(\mu/2)) + \delta_1}
{\epsilon_0(\gamma -\log(\mu/2)) + \delta_0} \mu^2\right. \\
&\quad\left. + \frac{1}{360}\left[\frac{3\left[\epsilon_2(\gamma -\log(\mu/2)) + \delta_2\right]}
{\epsilon_0(\gamma -\log(\mu/2)) + \delta_0}
- \frac{5\left[\epsilon_1(\gamma -\log(\mu/2)) + \delta_1\right]^2}
{\left[\epsilon_0(\gamma -\log(\mu/2)) + \delta_0\right]^2}\right]\mu^4
\right\}\,.
\end{aligned}
\end{equation}
The expansion coefficients $\epsilon_n(u)$ and $\delta_n(u)$ are defined in
\eqref{eq:e0_d0}--\eqref{eq:d2} in Appendix~\ref{sec:appendix_short_distance}
and depend only on the geometric parameter
\begin{equation}
\label{eq:def_u}
u = \frac{R_\mathrm{eff}^2}{R_1R_2}\,.
\end{equation}
This parameter can take arbitrary values between
$0$ and $1/4$, corresponding to the sphere-plane geometry and equally sized
spheres, respectively. Table~\ref{tab:short_distance_limits} gives the values
of the expansion coefficients for the two limiting cases. The sphere-plane limit, $u=0$,
is consistent with the results given in \cite{BimonteEmig2012}. Their numerical
constants $\gamma_i$, $i= 1, 2, 3, 4$ can now be expressed analytically as
\begin{align}
\gamma_1 &= \gamma + \log(2)\\
\gamma_2 &= \gamma_1 + \frac{1}{12}\\
\gamma_3 &= \frac{1}{2}\left(5\gamma_2^2 - 3\gamma_1^2 - \frac{107}{120}\gamma_1\right)\\
\gamma_4 &= 5\gamma_2 -3\gamma_1 - \frac{107}{240}\,.
\end{align}

\begin{table}
\centering
\begin{tabular}{cll}
$u$ & 0 &  $1/4$  \\
\hline
$\epsilon_0(u)$ & $1$ & $\log(2)$ \\
$\delta_0(u)$   & $0$ & $\log^2(2)$\\
$\epsilon_1(u)$ & $1$ & $\frac{1}{2}\left(\log(2) -\frac{1}{8}\right)$\\
$\delta_1(u)$ & $\frac{1}{12}$& $\frac{1}{2}\left(\log^2(2) -\frac{1}{12}\log(2)\right)$ \\
$\epsilon_2(u)$ & $1$ & $\frac{1}{3}\left(\log(2) - \frac{47}{128}\right)$ \\
$\delta_2(u)$ & $\frac{107}{360}$ & $\frac{1}{3}\left(\log^2(2) - \frac{83}{320}\log(2)
- \frac{5}{384}\right)$
\end{tabular} 
\caption{Expansion coefficients $\epsilon_n(u)$ and $\delta_n(u)$ appearing in
	\eqref{eq:short_distance} in the limits of the sphere-plane geometry
        ($u=0$) and of equal spheres ($u=1/4$).}
\label{tab:short_distance_limits}
\end{table}

\section{Conclusions}

We have for the first time derived an exact analytical expression for the Casimir free energy
of two Drude spheres of arbitrary radii completely within the scattering approach common in
Casimir physics. In contrast to previous work on the sphere-plane geometry and two spheres of
equal radii, the plane-wave basis was used, which led to a connection with a combinatorial
problem. The structure of this combinatorial problem highlights the difference between the
general two-sphere case and the corresponding limiting cases. The scattering approach also
provides an intuitive interpretation of the structure of the result for the Casimir free energy.

Earlier work by Fosco et al. \cite{Fosco2016} has pointed out the relevance of the capacitance
matrix for the Casimir free energy in the high-temperature limit. However, the fact that an
analytical expression for the capacitance matrix exists even for two spheres of different radii
seems to have largely escaped the attention of the Casimir community. Our work thus provides
an interesting connection between Casimir physics and electrostatics. This is in particular
the case for the short-distance expansion where by profiting from results obtained within the
electrostatics community, we derived a systematic expansion in powers of $\mu$ which might be
useful in that community as well.

\section*{Acknowledgements}
The authors are grateful to Michael Hartmann, Astrid Lambrecht, Paulo Maia Neto, Serge
Reynaud and Benjamin Spreng for many inspiring discussions. Benjamin Spreng has also kindly
provided numerical data for comparison between analytical and numerical results in the general
sphere-sphere geometry.

\begin{appendix}

\section{Coefficients of the short-distance expansion}
\label{sec:appendix_short_distance}

In this appendix, we derive expressions for the coefficients $\epsilon_0,
\delta_0, \epsilon_1, \delta_1, \epsilon_2$, and $\delta_2$ appearing in the
short-distance expansion \eqref{eq:short_distance}. While the expansion
parameter $\mu$ depends on the distance between the two spheres, the
geometric parameter $u$ defined in \eqref{eq:def_u} is a function of the sphere
radii alone. The parameter $v_0$ introduced in \eqref{eq:def_v0} can be
expressed in terms of $u$ as
\begin{equation}
\label{eq:def_v0_u}
v_{0} = \mathrm{sgn}(R_1 - R_2)\frac{\sqrt{1-4u}}{2}\,. 
\end{equation}
We remark that the sign of the difference of radii will not show up in the final
expressions. As we will start from the results \eqref{eq:taylor_Ic}--\eqref{eq:def_Jmc},
we need to express the coefficients $c_n$ in terms of the Taylor coefficients $v(\mu)$.
From $c=3/2\pm v$ and together with \eqref{eq:def_v0_u} we obtain for the first three
coefficients
\begin{align}
c_0 &= \frac{3}{2}\pm v_0\\
c_1 &= \mp\frac{v_0}{3}u\\
c_2 &= \mp\frac{v_0}{60}u(1-12u)\,.
\end{align}
For convenience of the reader, we list the partial ordinary Bell polynomials for the
index combinations needed in the following:
\begin{align}
\hat B_{n,0}(x_1, x_2,\ldots) &= \delta_{n, 0} \\
\hat B_{n,1}(x_1, x_2,\ldots) &= x_n \\
\hat B_{n,n-1}(x_1, x_2,\ldots) &= (n-1)x_1^{n-2}x_2 \\
\hat B_{n,n}(x_1, x_2,\ldots) &= x_1^n\,.
\end{align}

For the monopole contributions \eqref{eq:Delta_IJ} to order $\mu^4$ we need the
coefficients of $I(1), I^{(\pm)}$, and $J^{(\pm)}$ up to second order. After
some tedious but straightforward algebra, we obtain from \eqref{eq:expansion_Ic}
\begin{align}
I_0(1) &= \gamma - \log\left(\frac{\mu}{2}\right) \label{eq:app_I01}\\
I_1(1) &= \frac{1}{72} \\
I_2(1) &= \frac{7}{43200}\,. \label{eq:app_I21}
\end{align}
From \eqref{eq:def_I0c} and \eqref{eq:def_Imc}, we find
\begin{align}
I_0^{(\pm)} &= \gamma -\log\left(\frac{\mu}{2}\right) - \Psi_0^{(\pm)}  
\label{eq:app_I0pm}\\
I_1^{(\pm)} &= \frac{1}{72} - \frac{u}{12} + \frac{1}{6}\left(\frac{1}{2} \pm v_{0}\right)
	- \Psi^{(\pm)}_1
	\\
I_2^{(\pm)} &= \frac{7}{43200} + \frac{48u +73u^2}{1440} 
	- \frac{7+13u}{360}\left(\frac{1}{2} \pm v_{0}\right)
	-\Psi^{(\pm)}_2\,, \label{eq:app_I2pm}
\end{align}
where we introduced
\begin{align}
\label{eq:Psi0}
\Psi_0^{(\pm)} &= \psi\left(\frac{3}{2}\pm v_{0}\right) + \gamma \\
\label{eq:Psi1}
\Psi_1^{(\pm)} &= \pm v_{1}\psi^{(1)}\left(\frac{3}{2}\pm v_{0}\right) \\
\label{eq:Psi2}
\Psi_2^{(\pm)} &= \pm v_{2}\psi^{(1)}\left(\frac{3}{2}\pm v_{0}\right)
	          + \frac{v_{1}^2}{2}\psi^{(2)}\left(\frac{3}{2}\pm v_{0}\right)\,.
\end{align}
Including the Euler-Mascheroni constant in $\Psi_0^{(\pm)}$ will help to
simplify the final expressions. We note that the coefficients
\eqref{eq:app_I01}--\eqref{eq:app_I21} can be obtained from
\eqref{eq:app_I0pm}--\eqref{eq:app_I2pm} by choosing the lower sign and setting $v_0=1/2$, 
i.e. $u=0$. 

From \eqref{eq:def_Jmc}, we finally obtain
\begin{align}
J_0^{(\pm)} &=\frac{1}{2} \pm v_{0} \\
J_1^{(\pm)} &= \frac{1-3u}{6} \left(\frac{1}{2} \pm v_{0}\right) \\
J_2^{(\pm)} &= \frac{1-15u(1-3u)}{120} \left(\frac{1}{2} \pm v_{0}\right)\,.
\end{align}

We now insert the coefficients just derived into the monopole contributions
\eqref{eq:Delta_IJ} and sort the product terms by powers of $\mu$. The result
can be written as
\begin{equation}
\label{eq:short_distance_appendix}
\Delta = \frac{k_\mathrm{B}T}{2} \log\left\{
\sum_{n=0}^2 \frac{\mu^{2n}}{(2n+1)!}\left[\epsilon_n(u) \left(\gamma -\log\frac{\mu}{2}\right) 
+ \delta_n(u)\right]+ \mathcal{O}(\mu^6)\right\}\,,
\end{equation}
with the coefficients
\begin{align}
\label{eq:e0_d0}
\epsilon_{0}(u) &= 1-u\varphi_{0,0}\\
\delta_{0}(u) &= 1 - \varphi_{0,1} + u\theta_{0,0}
\end{align}
\begin{align}
\label{eq:e1}
\epsilon_1(u) &= 1- 2u -u^2 
	- 2u(1-3u)\varphi_{0,0} 
	- 6u\varphi_{1,0}\\
\label{eq:d1}
\delta_{1}(u) &= \frac{1}{12}(13-30u)
 			- \frac{1}{12}u(13-6u)\varphi_{0,0} 
 			- (1-4u)\varphi_{0,1}
 			- 6\varphi_{1,1} \\
&\quad 	+ 2u(1-3u)\theta_{0,0} 
		+ 6u(\theta_{0,1} + \theta_{1,0})
 \nonumber
\end{align}
\begin{align}
\label{eq:e2}
\epsilon_{2}(u) &= \frac{1}{6}(6- 64u +132u^2 + 193u^3)
	- \frac{2}{3}u(8-75u+180u^2)\varphi_{0,0}
\\&\quad
	- 40u(1-3u) \varphi_{1,0} - 120u\varphi_{2,0}
\nonumber\\
\label{eq:d2}
\delta_2(u) &= 
	\frac{1}{360}(467 - 5240u +14810u^2 +300u^3)
\\&\quad
	-\frac{1}{120}u(589 -3040u +1930 u^2)\varphi_{0,0}
	\nonumber
\\&\quad	
	-\frac{1}{3}(3-58u+208u^2)\varphi_{0,1}
	\nonumber
\\&\quad	
	- \frac{5}{3}u(13-6u)\varphi_{1,0} - 20(1-4u)\varphi_{1,1}
	-120 \varphi_{2,1}
	\nonumber
\\&\quad	
	+\frac{2}{3}u(8-75u+180u^2)\theta_{0,0}
	\nonumber
\\&\quad		
	+ 40 u(1-3u)(\theta_{0,1} + \theta_{1,0})
	+ 120 u\theta_{1,1}
	+ 120 u (\theta_{0,2} + \theta_{2,0}) \,.
	\nonumber
\end{align}
Here, we have introduced abbreviations for the sums of the functions
\eqref{eq:Psi0}--\eqref{eq:Psi2}
\begin{equation}
\varphi_{n,m} = \left(\frac{1}{2}+v_{0}\right)^m\Psi^{(+)}_n +
\left(\frac{1}{2}-v_{0}\right)^m\Psi^{(-)}_n
\end{equation}
as well as for their products
\begin{equation}
\theta_{n,m} = \Psi^{(+)}_n \Psi^{(-)}_m\,. 
\end{equation}

For the limiting cases of the sphere-plane geometry ($u=0$) and two
spheres of equal radii ($u=1/4$), the coefficients $v_{1}$ and 
$v_{2}$ vanish. Hence, $\varphi_{1,m}$ and $\varphi_{2,m}$ yield zero 
and all $\theta_{n,m}$ except for $\theta_{0,0}$ vanish. 
The expressions for $\epsilon_n$ and $\delta_n$ then simplify
to the results listed in Table~\ref{tab:short_distance_limits}.
\end{appendix}

\bibliography{hightemperature}

\begin{thebibliography}{10}
\providecommand{\url}[1]{\texttt{#1}}
\providecommand{\urlprefix}{URL }
\expandafter\ifx\csname urlstyle\endcsname\relax
  \providecommand{\doi}[1]{doi:\discretionary{}{}{}#1}\else
  \providecommand{\doi}{doi:\discretionary{}{}{}\begingroup
  \urlstyle{rm}\Url}\fi
\providecommand{\eprint}[2][]{\url{#2}}

\bibitem{Feinberg2001}
J.~Feinberg, A.~Mann and M.~Revzen,
\newblock \emph{Casimir {Effect}: {The} {Classical} {Limit}},
\newblock Ann. Phys. (N.Y.) \textbf{288}, 103 (2001),
\newblock \doi{10.1006/aphy.2000.6118}.

\bibitem{LambrechtNetoReynaud2006}
A.~Lambrecht, P.~A. Maia~Neto and S.~Reynaud,
\newblock \emph{The {Casimir} effect within scattering theory},
\newblock New J. Phys. \textbf{8}, 243 (2006),
\newblock \doi{10.1088/1367-2630/8/10/243}.

\bibitem{Sauer1962}
F.~Sauer,
\newblock \emph{Die {Temperaturabhängigkeit} von {Dispersionskräften}},
\newblock Ph.D. thesis, Universität Göttingen (1962).

\bibitem{Mehra1967}
J.~Mehra,
\newblock \emph{Temperature correction to the {Casimir} effect},
\newblock Physica \textbf{37}, 145 (1967),
\newblock \doi{10.1016/0031-8914(67)90115-2}.

\bibitem{BimonteEmig2012}
G.~Bimonte and T.~Emig,
\newblock \emph{Exact {Results} for {Classical} {Casimir} {Interactions}:
  {Dirichlet} and {Drude} {Model} in the {Sphere}-{Sphere} and {Sphere}-{Plane}
  {Geometry}},
\newblock Phys. Rev. Lett. \textbf{109}, 160403 (2012),
\newblock \doi{10.1103/PhysRevLett.109.160403}.

\bibitem{Bimonte2018}
G.~Bimonte,
\newblock \emph{Beyond-proximity-force-approximation {Casimir} force between
  two spheres at finite temperature},
\newblock Phys. Rev. D \textbf{97}, 085011 (2018),
\newblock \doi{10.1103/PhysRevD.97.085011}.

\bibitem{Ether2015}
{D. S. Ether jr., L. B. Pires, S. Umrath, D. Martinez, Y. Ayala, B. Pontes, G.
  R. de S. Araújo, S. Frases, G.-L. Ingold, F. S. S. Rosa, N. B. Viana, H. M.
  Nussenzveig and P. A. Maia Neto},
\newblock \emph{Probing the {Casimir} force with optical tweezers},
\newblock EPL \textbf{112}, 44001 (2015),
\newblock \doi{10.1209/0295-5075/112/44001}.

\bibitem{Garrett2018}
J.~L. Garrett, D.~A.~T. Somers and J.~N. Munday,
\newblock \emph{Measurement of the {Casimir} {Force} between {Two} {Spheres}},
\newblock Phys. Rev. Lett. \textbf{120}, 040401 (2018),
\newblock \doi{10.1103/PhysRevLett.120.040401}.

\bibitem{Elzbieciak-Wodka2014}
M.~Elzbieciak-Wodka, M.~N. Popescu, F.~J.~M. Ruiz-Cabello, G.~Trefalt,
  P.~Maroni and M.~Borkovec,
\newblock \emph{Measurements of dispersion forces between colloidal latex
  particles with the atomic force microscope and comparison with {Lifshitz}
  theory},
\newblock J. Chem. Phys. \textbf{140}, 104906 (2014),
\newblock \doi{10.1063/1.4867541}.

\bibitem{Ruiz-Cabello2017}
F.~J.~M. Ruiz-Cabello, M.~Moazzami-Gudarzi, M.~Elzbieciak-Wodka and P.~Maroni,
\newblock \emph{{Forces Between Different Latex Particles in Aqueous
  Electrolyte Solutions Measured with the Colloidal Probe Technique}},
\newblock Microsc. Res. Tech. \textbf{80}, 144 (2017),
\newblock \doi{10.1002/jemt.22656}.

\bibitem{Zhao2013}
R.~Zhao, Y.~Luo, A.~I. Fernández-Domínguez and J.~B. Pendry,
\newblock \emph{Description of van der {Waals} {Interactions} {Using}
  {Transformation} {Optics}},
\newblock Phys. Rev. Lett. \textbf{111}, 033602 (2013),
\newblock \doi{10.1103/PhysRevLett.111.033602}.

\bibitem{Fosco2016}
C.~Fosco, F.~Lombardo and F.~Mazzitelli,
\newblock \emph{Casimir free energy at high temperatures: {Grounded} versus
  isolated conductors},
\newblock Phys. Rev. D \textbf{93}, 125015 (2016),
\newblock \doi{10.1103/PhysRevD.93.125015}.

\bibitem{Spreng2018}
B.~Spreng, M.~Hartmann, V.~Henning, P.~A. Maia~Neto and G.-L. Ingold,
\newblock \emph{Proximity force approximation and specular reflection:
  {Application} of the {WKB} limit of {Mie} scattering to the {Casimir}
  effect},
\newblock Phys. Rev. A \textbf{97}, 062504 (2018),
\newblock \doi{10.1103/PhysRevA.97.062504}.

\bibitem{Henning2019}
V.~Henning, B.~Spreng, M.~Hartmann, G.-L. Ingold and P.~A. Maia~Neto,
\newblock \emph{Role of diffraction in the {Casimir} effect beyond the
  proximity force approximation},
\newblock J. Opt. Soc. Am. B \textbf{36}, C77 (2019),
\newblock \doi{10.1364/JOSAB.36.000C77}.

\bibitem{Spreng2020}
B.~Spreng, P.~A. Maia~Neto and G.-L. Ingold,
\newblock \emph{Plane-wave approach to the exact van der {Waals} interaction
  between colloid particles},
\newblock J. Chem. Phys. \textbf{153}, 024115 (2020),
\newblock \doi{10.1063/5.0011368}.

\bibitem{Smythe1950}
W.~C. Smythe,
\newblock \emph{Static and {Dynamic} {Electricity}},
\newblock McGraw-Hill, New York,
\newblock ISBN 978-0891169178,
\newblock Chap. 5 (1950).

\bibitem{CanaguierDurand2010}
A.~Canaguier-Durand, P.~A. Maia~Neto, A.~Lambrecht and S.~Reynaud,
\newblock \emph{Thermal {Casimir} effect for {Drude} metals in the plane-sphere
  geometry},
\newblock Phys. Rev. A \textbf{82}, 012511 (2010),
\newblock \doi{10.1103/PhysRevA.82.012511}.

\bibitem{MaiaNeto2008}
P.~A. Maia~Neto, A.~Lambrecht and S.~Reynaud,
\newblock \emph{Casimir energy between a plane and a sphere in electromagnetic
  vacuum},
\newblock Phys. Rev. A \textbf{78}, 012115 (2008),
\newblock \doi{10.1103/PhysRevA.78.012115}.

\bibitem{Emig2008}
T.~Emig,
\newblock \emph{Fluctuation-induced quantum interactions between compact
  objects and a plane mirror},
\newblock J. Stat. Mech. \textbf{2008}, P04007 (2008),
\newblock \doi{10.1088/1742-5468/2008/04/p04007}.

\bibitem{Nieto-Vesperinas2006}
M.~Nieto-Vesperinas,
\newblock \emph{Scattering and {Diffraction} in {Physical} {Optics}},
\newblock World Scientific, Singapore,
\newblock ISBN 978-981-256-340-8,
\newblock \doi{10.1142/5833} (2006).

\bibitem{BohrenHuffman2004}
C.~F. Bohren and D.~R. Huffman,
\newblock \emph{Absorption and {Scattering} of {Light} by {Small} {Particles}},
\newblock Wiley-VCH, Weinheim,
\newblock ISBN 978-0-471-29340-8,
\newblock \doi{10.1002/9783527618156},
\newblock Chap. 4 (2004).

\bibitem{Molinari1997}
L.~Molinari,
\newblock \emph{Transfer matrices and tridiagonal-block {Hamiltonians} with
  periodic and scattering boundary conditions},
\newblock J. Phys. A: Math. Gen. \textbf{30}, 983 (1997),
\newblock \doi{10.1088/0305-4470/30/3/021}.

\bibitem{Molinari2008}
L.~G. Molinari,
\newblock \emph{Determinants of block tridiagonal matrices},
\newblock Linear Algebra Appl. \textbf{429}, 2221 (2008),
\newblock \doi{10.1016/j.laa.2008.06.015}.

\bibitem{GoverBarnett1985}
M.~J.~C. Gover and S.~Barnett,
\newblock \emph{Inversion of {Toeplitz} {Matrices} which are not {Strongly}
  {Non}-singular},
\newblock IMA J Numer Anal \textbf{5}, 101 (1985),
\newblock \doi{10.1093/imanum/5.1.101}.

\bibitem{Gover1994}
M.~J.~C. Gover,
\newblock \emph{The eigenproblem of a tridiagonal 2-{Toeplitz} matrix},
\newblock Linear Algebra Appl. \textbf{197-198}, 63 (1994),
\newblock \doi{10.1016/0024-3795(94)90481-2}.

\bibitem{Comtet1974}
L.~Comtet,
\newblock \emph{Advanced combinatorics},
\newblock D. Reidel, Dordrecht,
\newblock ISBN 90-277-0380-9 (1974).

\bibitem{Maxwell1873}
J.~C. Maxwell,
\newblock \emph{A {Treatise} on {Electricity} and {Magnetism}},
\newblock Clarendon Press., Oxford,
\newblock \S 173 (1873).

\bibitem{BalianDuplantier1978}
R.~Balian and B.~Duplantier,
\newblock \emph{Electromagnetic waves near perfect conductors. {II}. {Casimir}
  effect},
\newblock Ann. Phys. (N.Y.) \textbf{112}, 165 (1978),
\newblock \doi{10.1016/0003-4916(78)90083-0}.

\bibitem{Garvin1936}
M.~C. Garvin,
\newblock \emph{A {Generalized} {Lambert} {Series}},
\newblock Am. J. Math. \textbf{58}, 507 (1936),
\newblock \doi{10.2307/2370967}.

\bibitem{BanerjeeWilkerson2017}
S.~Banerjee and B.~Wilkerson,
\newblock \emph{Asymptotic expansions of {Lambert} series and related
  q-series},
\newblock Int. J. Number Theory \textbf{13}, 2097 (2017),
\newblock \doi{10.1142/S1793042117501135}.

\bibitem{dlmf}
\emph{{NIST Digital Library of Mathematical Functions}},
\newblock http://dlmf.nist.gov/, Release 1.0.28 of 2020-09-15,
\newblock F.~W.~J. Olver, A.~B. {Olde Daalhuis}, D.~W. Lozier, B.~I. Schneider,
  R.~F. Boisvert, C.~W. Clark, B.~R. Miller, B.~V. Saunders, H.~S. Cohl, and
  M.~A. McClain, eds.

\bibitem{Lekner2011}
J.~Lekner,
\newblock \emph{Capacitance coefficients of two spheres},
\newblock J. Electrostat. \textbf{69}, 11 (2011),
\newblock \doi{10.1016/j.elstat.2010.10.002}.

\bibitem{Banerjee2019}
S.~Banerjee, T.~Peters, Y.~Song and B.~Wilkerson,
\newblock \emph{Closed-form and asymptotic capacitance coefficients for the
  electrostatics of two spheres},
\newblock J. Electrostat. \textbf{101}, 103369 (2019),
\newblock \doi{10.1016/j.elstat.2019.103369}.

\end{thebibliography}

\end{document}